\documentclass[twoside,twocolumn,9pt]{article}
\usepackage{extsizes}
\usepackage[super,sort&compress,comma]{natbib} 
\usepackage[version=3]{mhchem}
\usepackage[left=1.5cm, right=1.5cm, top=1.785cm, bottom=2.0cm]{geometry}
\usepackage{packages/balance}
\usepackage{packages/widetext}
\usepackage{packages/times,packages/mathptmx}
\usepackage{sectsty}
\usepackage{graphicx} 
\usepackage{lastpage}
\usepackage[format=plain,justification=raggedright,singlelinecheck=false,font={stretch=1.125,small,sf},labelfont=bf,labelsep=space]{caption}
\usepackage{float}
\usepackage{fancyhdr}
\usepackage{fnpos}
\usepackage[english]{babel}
\usepackage{array}
\usepackage{droidsans}
\usepackage{charter}
\usepackage[T1]{fontenc}
\usepackage[usenames,dvipsnames]{xcolor}
\usepackage{setspace}
\usepackage[compact]{titlesec}

\usepackage[normalem]{ulem}

\usepackage{psfrag}

\usepackage{epstopdf}%This line makes .eps figures into .pdf - please comment out if not required.

\definecolor{cream}{RGB}{222,217,201}

\begin{document}

\pagestyle{fancy}
\thispagestyle{plain}
\fancypagestyle{plain}{

%%%HEADER%%%
\fancyhead[C]{\includegraphics[width=18.5cm]{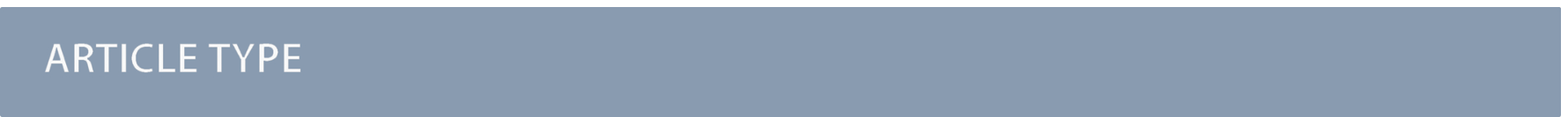}}
\fancyhead[L]{\hspace{0cm}\vspace{1.5cm}\includegraphics[height=30pt]{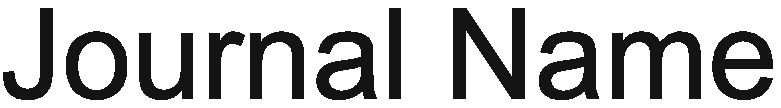}}
\fancyhead[R]{\hspace{0cm}\vspace{1.7cm}\includegraphics[height=55pt]{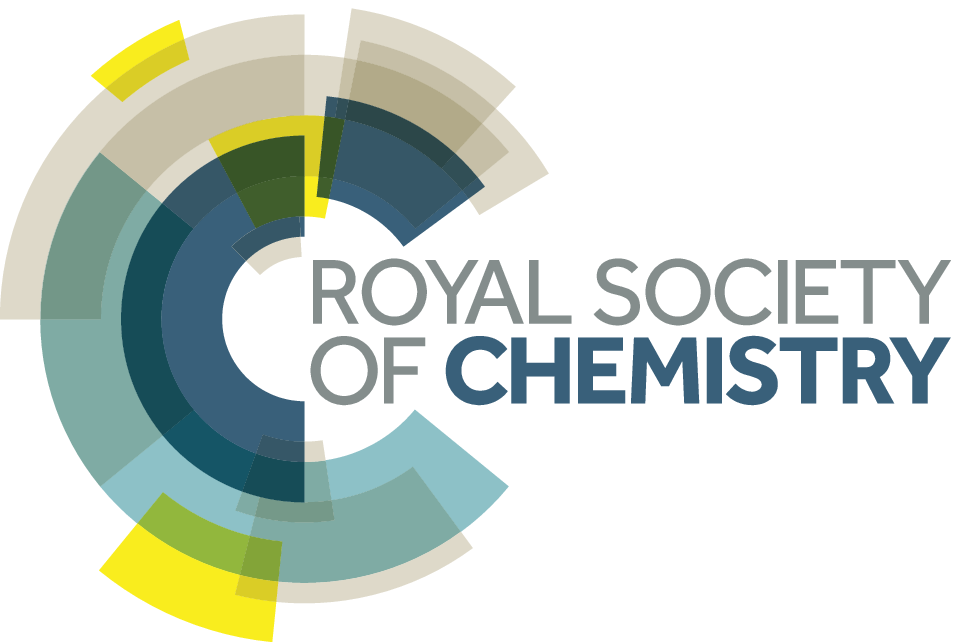}}
\renewcommand{\headrulewidth}{0pt}
}
%%%END OF HEADER%%%

%%%PAGE SETUP - Please do not change any commands within this section%%%
\makeFNbottom
\makeatletter
\renewcommand\LARGE{\@setfontsize\LARGE{15pt}{17}}
\renewcommand\Large{\@setfontsize\Large{12pt}{14}}
\renewcommand\large{\@setfontsize\large{10pt}{12}}
\renewcommand\footnotesize{\@setfontsize\footnotesize{7pt}{10}}
\makeatother

\renewcommand{\thefootnote}{\fnsymbol{footnote}}
\renewcommand\footnoterule{\vspace*{1pt}% 
\color{cream}\hrule width 3.5in height 0.4pt \color{black}\vspace*{5pt}} 
\setcounter{secnumdepth}{5}

\makeatletter 
\renewcommand\@biblabel[1]{#1}            
\renewcommand\@makefntext[1]% 
{\noindent\makebox[0pt][r]{\@thefnmark\,}#1}
\makeatother 
\renewcommand{\figurename}{\small{Fig.}~}
\sectionfont{\sffamily\Large}
\subsectionfont{\normalsize}
\subsubsectionfont{\bf}
\setstretch{1.125} %In particular, please do not alter this line.
\setlength{\skip\footins}{0.8cm}
\setlength{\footnotesep}{0.25cm}
\setlength{\jot}{10pt}
\titlespacing*{\section}{0pt}{4pt}{4pt}
\titlespacing*{\subsection}{0pt}{15pt}{1pt}
%%%END OF PAGE SETUP%%%

%%%FOOTER%%%
\fancyfoot{}
\fancyfoot[LO,RE]{\vspace{-7.1pt}\includegraphics[height=9pt]{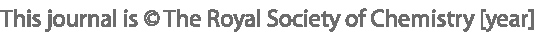}}
\fancyfoot[CO]{\vspace{-7.1pt}\hspace{13.2cm}\includegraphics{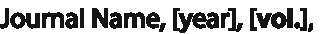}}
\fancyfoot[CE]{\vspace{-7.2pt}\hspace{-14.2cm}\includegraphics{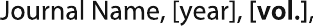}}
\fancyfoot[RO]{\footnotesize{\sffamily{1--\pageref{LastPage} ~\textbar  \hspace{2pt}\thepage}}}
\fancyfoot[LE]{\footnotesize{\sffamily{\thepage~\textbar\hspace{3.45cm} 1--\pageref{LastPage}}}}
\fancyhead{}
\renewcommand{\headrulewidth}{0pt} 
\renewcommand{\footrulewidth}{0pt}
\setlength{\arrayrulewidth}{1pt}
\setlength{\columnsep}{6.5mm}
\setlength\bibsep{1pt}
%%%END OF FOOTER%%%

%%%FIGURE SETUP - please do not change any commands within this section%%%
\makeatletter 
\newlength{\figrulesep} 
\setlength{\figrulesep}{0.5\textfloatsep} 

\newcommand{\topfigrule}{\vspace*{-1pt}% 
\noindent{\color{cream}\rule[-\figrulesep]{\columnwidth}{1.5pt}} }

\newcommand{\botfigrule}{\vspace*{-2pt}% 
\noindent{\color{cream}\rule[\figrulesep]{\columnwidth}{1.5pt}} }

\newcommand{\dblfigrule}{\vspace*{-1pt}% 
\noindent{\color{cream}\rule[-\figrulesep]{\textwidth}{1.5pt}} }

\makeatother
%%%END OF FIGURE SETUP%%%

\newcommand{\xm}{x_{\rm m}}

%%%TITLE, AUTHORS AND ABSTRACT%%%
\twocolumn[
  \begin{@twocolumnfalse}
\vspace{3cm}
\sffamily
\begin{tabular}{m{4.5cm} p{13.5cm} }

\includegraphics{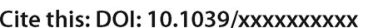} & \noindent\LARGE{\textbf{Using evaporation to control capillary instabilities in micro-systems$^\dag$}} \\%Article title goes here instead of the text "This is the title"
\vspace{0.3cm} & \vspace{0.3cm} \\

 & \noindent\large{Rodrigo Ledesma-Aguilar,$^{\ast}$\textit{$^{a}$} Gianluca Laghezza,\textit{$^{b}$} Julia M. Yeomans,\textit{$^{b}$} and Dominic Vella\textit{$^{c}$}} \\%Author names go here instead of "Full name", etc.

\includegraphics{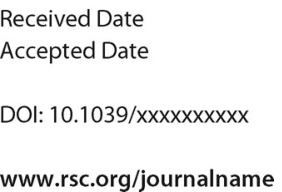} & \noindent\normalsize{
The instabilities of fluid interfaces represent both a limitation and an opportunity for the fabrication of small-scale devices. 
Just as  non-uniform capillary pressures can destroy micro-electrical mechanical systems (MEMS), 
so they can guide the assembly of novel solid and fluid structures. 
In many such applications the interface appears during an evaporation process and is therefore only present temporarily. 
It is commonly assumed that this evaporation simply guides the interface through a sequence of equilibrium configurations, 
and that the rate of evaporation only sets the timescale of this sequence. 
Here, we use Lattice-Boltzmann simulations and a theoretical analysis 
to show that, in fact, the rate of evaporation can be a factor in determining the onset and form of dynamical capillary instabilities. 
Our results shed light on the role of evaporation in previous experiments, 
and open the possibility of exploiting diffusive mass transfer to directly control capillary flows in MEMS applications.
} \\%The abstrast goes here instead of the text "The abstract should be..."

\end{tabular}

 \end{@twocolumnfalse} \vspace{0.6cm}

  ]
%%%END OF TITLE, AUTHORS AND ABSTRACT%%%

%%%FONT SETUP - please do not change any commands within this section
\renewcommand*\rmdefault{bch}\normalfont\upshape
\rmfamily
\section*{}
\vspace{-1cm}

%%%FOOTNOTES%%%

\footnotetext{\textit{$^{a}$~Smart Materials and Surfaces Laboratory, Northumbria University, Ellison Place,  Newcastle upon Tyne NE1 8ST, United Kingdom; E-mail: rodrigo.ledesma@northumbria.ac.uk}}
\footnotetext{\textit{$^{b}$~The Rudolf Peierls Centre for Theoretical Physics, University of Oxford, 1 Keble Road, Oxford OX1 3NP, United Kingdom}}
\footnotetext{\textit{$^{c}$~Mathematical Institute, Andrew Wiles Building, University of Oxford, Woodstock Road, Oxford OX2 6GG, United Kingdom}}

%Please use \dag to cite the ESI in the main text of the article.
%If you article does not have ESI please remove the the \dag symbol from the title and the footnotetext below.
\footnotetext{\dag~Electronic Supplementary Information (ESI) available: Details of lattice-Boltzmann numerical simulations. See DOI: 10.1039/b000000x/}
%additional addresses can be cited as above using the lower-case letters, c, d, e... If all authors are from the same address, no letter is required

%\footnotetext{\ddag~Additional footnotes to the title and authors can be included \emph{e.g.}\ `Present address:' or `These authors contributed equally to this work' as above using the symbols: \ddag, \textsection, and \P. Please place the appropriate symbol next to the author's name and include a \texttt{\textbackslash footnotetext} entry in the the correct place in the list.}

%%%END OF FOOTNOTES%%%

%%%MAIN TEXT%%%%

\section{Introduction}

Two-phase flows at small scales occur in many systems from flow in porous media to microfluidics and the intermediate stages of micro-electrical mechanical systems (MEMS) fabrication.  
At these small scales, the interface between two fluids, or between a fluid and its vapour, is particularly relevant, since the forces from surface tension play a dominant role. 
The form taken by the interface is also very important: because capillary flows arise from deformations of the shape of an interface, the presence of a morphological instability can amplify 
the effect of surface tension forces. 
Such instabilities are notorious for their destructive potential in  MEMS fabrication: surface tension forces often cause permanent damage to very small components, which tend to be more flexible because of the scales involved \cite{Tanaka1993,Abe1995,Abe1996}.

While the traditional point of view has been that capillary instabilities at micro-scales should be understood so that they can be avoided, recently it has become apparent that various features of  such instabilities may also be useful. 
For example, unstable fronts can be used to form regular arrays in carbon nanotube forests \cite{Chakrapani2004,deVolder2013a,deVolder2013b}, mesoscale bristles \cite{Pokroy2009} and to displace a liquid trapped within a confined region \cite{Keiser2016}. 
In these situations, the question is then not how to avoid an instability so much as how to control it. 
A similar point of view has recently emerged in solid mechanics, where buckling instabilities are being actively revisited due to their utility in a range of applications \cite{Reis2015}. 

In controlling the onset and development of capillary instabilities, it is well known that the geometry and wettability of the system can be extremely important. 
For example, the tapering of a two-dimensional channel has been shown to stabilize interfaces against the Saffman-Taylor instability~\cite{AlHousseiny2012,al2013two}, 
while such tapering can also introduce an interfacial instability  allowing water to displace oil confined in the apex ~\cite{Keiser2016}.  
Tapering can also be an important component in microscopic systems as a means to move liquids spontaneously~\cite{Reyssat2014,Gorce2016}. 
Similarly, gradients in wettability,  caused by chemical patterning~\cite{chandesris2013uphill} or thermo-capillary (Marangoni) effects~\cite{kataoka1999patterning}, can move liquids on solid surfaces and can also give rise to instability~\cite{konnur2000instability,honisch2015instabilities}.

A common feature in many micro-scale systems is evaporation. 
For example, in the `densification' of carbon nanotubes \cite{deVolder2013a,deVolder2013b} a liquid is required to cause the desired instability, but is also removed during the process. 
Similarly, in MEMS fabrication an interface is only introduced during the drying of a rinsing agent \cite{Raccurt2004}. 
Despite the importance of evaporation in these processes, the role of the \emph{rate} of evaporation has been neglected, being assumed merely to give a time scale to the process of interest. 
Nevertheless, there is anecdotal evidence that this might not be the sole effect of the evaporation rate in such applications. 
For example, in the creation of nanotube foams, Chakrapani \emph{et al.} \cite{Chakrapani2004} report that the cell size within the foam depends on the relative humidity (and hence on the rate of evaporation). 

A previous reduced model for the role of evaporation rate~\cite{Hadjittofis2016} was motivated by the observed dependence of final patterns seen in dense carbon nanotube foams on the evaporation rate \cite{Chakrapani2004}.  This model accounted for the liquid films that form between deformable elastic elements so that, while surface tension acts to bring them together in an instability, elasticity as well as lubrication forces act to separate them. While Hadjittofis \emph{et al.} \cite{Hadjittofis2016} were able to explain experimental observations qualitatively, they could do so only if the evaporation rate was extremely large. Essentially, the theoretical model showed that for sufficiently large evaporation rates the liquid phase would evaporate before the deformed objects could squeeze out the liquid film that separated them.  
While this result is perhaps intuitively obvious, it is difficult to imagine this being the key mechanism in the observed role of evaporation rate. 

\begin{figure}[b!]
\centering
\psfrag{a}{(a)}
\psfrag{b}{(b)}
\psfrag{T}{Time}
\includegraphics[width=0.45\textwidth]{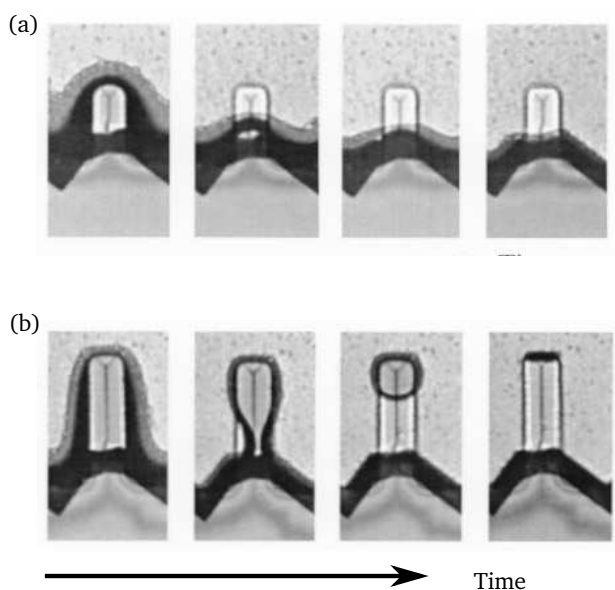}
\caption{\label{fig:Abe} {Top-view experimental images showing the evaporation of a rinse liquid from cantilever structures (Copyright 1995 IEEE. Reprinted, with permission, from Abe \& Reed, \emph{J. Microelectromech. Syst.} \textbf{4}, 66--75 (1995)). (a) On a short (and hence relatively rigid) cantilever, relatively little bending due to surface tension occurs and the liquid front remains stable upon evaporation. (b) On a longer (and hence relatively flexible) lever, surface tension forces create a more significant tapering, and the retraction of the wetting front leads to instability and the pinch-off of a drop. }}
\end{figure}

{The possibility that new hydrodynamic mechanisms of instability affected by evaporation might emerge is, in part, motivated by the experimental observations of \citet{Abe1995}; \citet{Abe1996}. 
In their experiments, evaporation of a rinse liquid around slender structures of different length led either to the smooth retraction of the meniscus around the  structure (see Fig.~\ref{fig:Abe}(a)) or to a pinch-off instability, with a droplet of the evaporating liquid temporarily isolated at the end of the structure (see Fig.~\ref{fig:Abe}(b)). 
The qualitative difference between the evaporation modes of Fig.~\ref{fig:Abe}(a) and \ref{fig:Abe}(b) is thought to be due to the difference in tapering between long (and hence flexible) structures, compared with shorter (and hence relatively rigid) structures. 
The mechanism behind drop pinch-off is essentially that discussed in recent studies of the stability of planar fluid fronts in tapered geometries \cite{AlHousseiny2012,al2013two}: surface tension pulls on the structure, creating a tapered channel where differences in capillary pressure drive pinch-off. 
Whether such an instability sets in is significant for the ultimate stiction of the structure because the deflection is imagined to be larger when a droplet can pull on a longer `lever arm'.
\citet{Abe1996} reported that structures are less prone to `stiction' if the rinsing and drying steps are performed at a higher temperature. 
The magnitude of the observed effect could not be rationalized using the changes in liquid properties that occur at higher temperatures \cite{Abe1995}, and  so it is natural to wonder whether 
instead it is the evaporation rate that matters?

In this paper we investigate whether qualitatively new mechanisms might emerge as the evaporation rate changes relative to the background effects of geometry and wettability; 
such mechanisms might explain how even moderate changes in the evaporation rate can effect a  qualitative change in the form of capillary instability that is observed experimentally. 
We consider the dynamics of evaporation in a geometry analogous to the MEMS setup of \citet{Abe1995}; \citet{Abe1996}:  a liquid occupies the channel formed beneath a solid cantilever~(see Fig.~\ref{fig:Sketch3D}).  While our geometry is motivated by specific experiments, it is representative of a range of MEMS and microfluidic applications.  We study the dynamics of the evaporating front using a Lattice--Boltzmann algorithm capable of modelling evaporation in contact with solid surfaces~\cite{Ledesma2014}.} We focus on isothermal situations, which correspond to the limit of a high thermal conductivity in the liquid~\cite{Cazabat2010}, and neglect Marangoni effects for simplicity.
We consider the effects of channel tapering and gradients in wettability together with the  evaporative driving (or the rate of evaporation). 
We investigate the conditions under which the two modes of instability reported experimentally (see Fig.~\ref{fig:Abe}) are observed. In particular, we show that,
as the evaporation rate alone varies the  pinch-off instability may be tiggered or suppressed;
{sufficient levels of evaporation suppress this instability since diffusion enhances the evaporative mass flux from a perturbation, smoothing that perturbation out. Crucially, a hydrodynamic instability driven by gradients in geometry or wettability (through which the interface ceases to be planar and becomes convoluted) may be suppressed by evaporation.}
To understand the emergence of this mechanism we present a linear stability analysis of an interface in a simplified geometry. 
This analysis reveals that it is the combination of evaporation rate, channel geometry and wettability that controls the stability of the liquid front. 

\section{Results}

\subsection{Lattice-Boltzmann simulations}

\label{sec:Simulations}

\begin{figure}[t!]
\psfrag{T}[r][r][1][0]{Top wall}
\psfrag{B}{Back wall}
\psfrag{M}{Bottom wall}
\psfrag{H}{$H$}
\psfrag{L}{$L$}
\psfrag{l}{$l$}
\psfrag{h}{$h+\Delta h$}
\psfrag{h1}{$h$}
\psfrag{W}{$W$}
\psfrag{w}{$w$}
\psfrag{p}{$\phi = \phi_0$}
\psfrag{p}{$S$}
\psfrag{a}{(a)}
\psfrag{b}{(b)}
\psfrag{Q}{Fluid}
\includegraphics[width=\columnwidth]{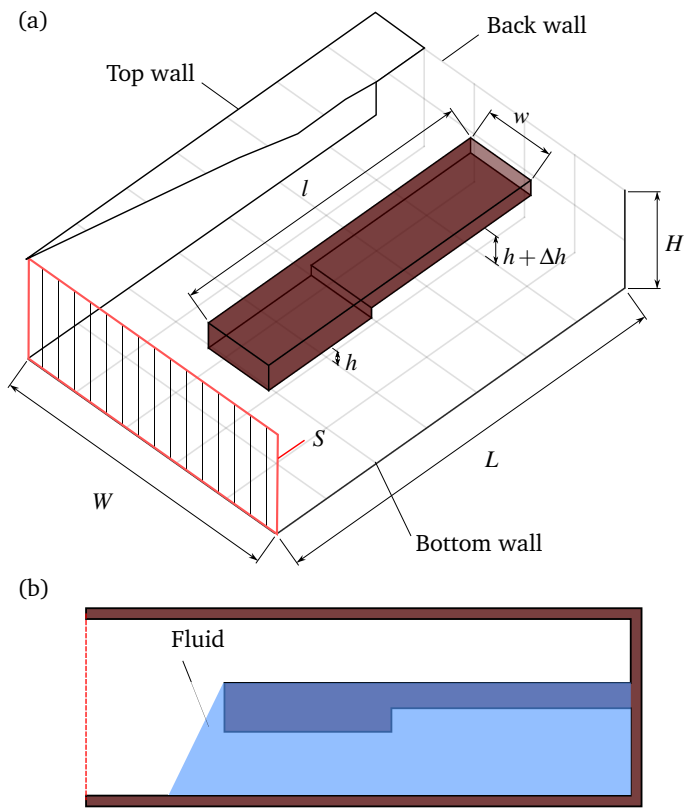}
\caption{\label{fig:Sketch3D}Schematic representation of the simulation domain. (a) A cantilever is fixed to a C-shaped micro cavity. An open boundary plane at the front of the cavity ($S$) drives the evaporation of the liquid by maintaining the local vapour concentration below the saturation point. (b) Side view of the system geometry. A fluid phase initially wets the cantilever structure before evaporating.}
\end{figure}

To illustrate the effect of the rate of evaporation on fluid dynamics in MEMS devices, we carried out numerical simulations of the evaporation of a liquid within a micro-etched structure using a lattice-Boltzmann algorithm~\cite{Ledesma2014} (see Supporting Information section for details). 
Fig.~\ref{fig:Sketch3D}(a) shows a schematic of the simulation geometry, which consists of a rectangular micro-beam fixed to the wall of a microfluidic cavity. 
{
The cavity is a C-shape channel formed by a back wall of height $H$ that joins two horizontal walls of length $L$ (marked top and bottom walls in the Fig.~\ref{fig:Sketch3D}). 
The beam, of length $l$, is attached to the back wall, and forms an overhanging cantilever of width $w$, suspended above the floor of the cavity at a local height $h$. 
The height step, $\Delta h$, allows us to study situations where there is a change in confinement between the cantilever and the bottom wall. 
In the following, we refer to the resulting structure as a stepped cantilever.
} 
We use periodic boundary conditions (so that in effect the structure is repeated in the transverse direction with a separation distance between the centres of the cantilevers, $W$). 
This geometry is typical of applications in microfluidics and MEMS. 

\begin{figure*}[!t]
\centering
 \psfrag{t1}[l][l]{$\tilde t = 0.01$}
 \psfrag{t2}[l][l]{$\tilde t = 0.15$}
 \psfrag{t3}[l][l]{$\tilde t = 0.29$}
 \psfrag{t4}[l][l]{$\tilde t = 0.44$}
 \psfrag{t5}[l][l]{$\tilde t = 0.51$}
 \psfrag{t6}[l][l]{$\tilde t = 0.58$}
 \psfrag{p1}[l][l]{$\tilde t = 0.01$}
 \psfrag{p2}[l][l]{$\tilde t = 0.15$}
 \psfrag{p3}[l][l]{$\tilde t = 0.29$}
 \psfrag{p4}[l][l]{$\tilde t = 0.44$}
 \psfrag{p5}[l][l]{$\tilde t = 0.45$}
 \psfrag{p6}[l][l]{$\tilde t = 0.47$}
 \psfrag{q1}[l][l]{$\tilde t = 0.01$}
\psfrag{q2}[l][l]{$\tilde t = 0.10$}
\psfrag{q3}[l][l]{$\tilde t = 0.19$}
\psfrag{q4}[l][l]{$\tilde t = 0.29$}
\psfrag{q5}[l][l]{$\tilde t = 0.31$}
\psfrag{q6}[l][l]{$\tilde t = 0.33$}
 \psfrag{A}[l][l]{(a) Uniform lever, ${\rm E}=0.020$}
 \psfrag{B}[l][l]{(b) Stepped lever, ${\rm E}=0.020$}
  \psfrag{C}[l][l]{(c) Stepped lever, ${\rm E}=0.027$}
\includegraphics[width=\textwidth]{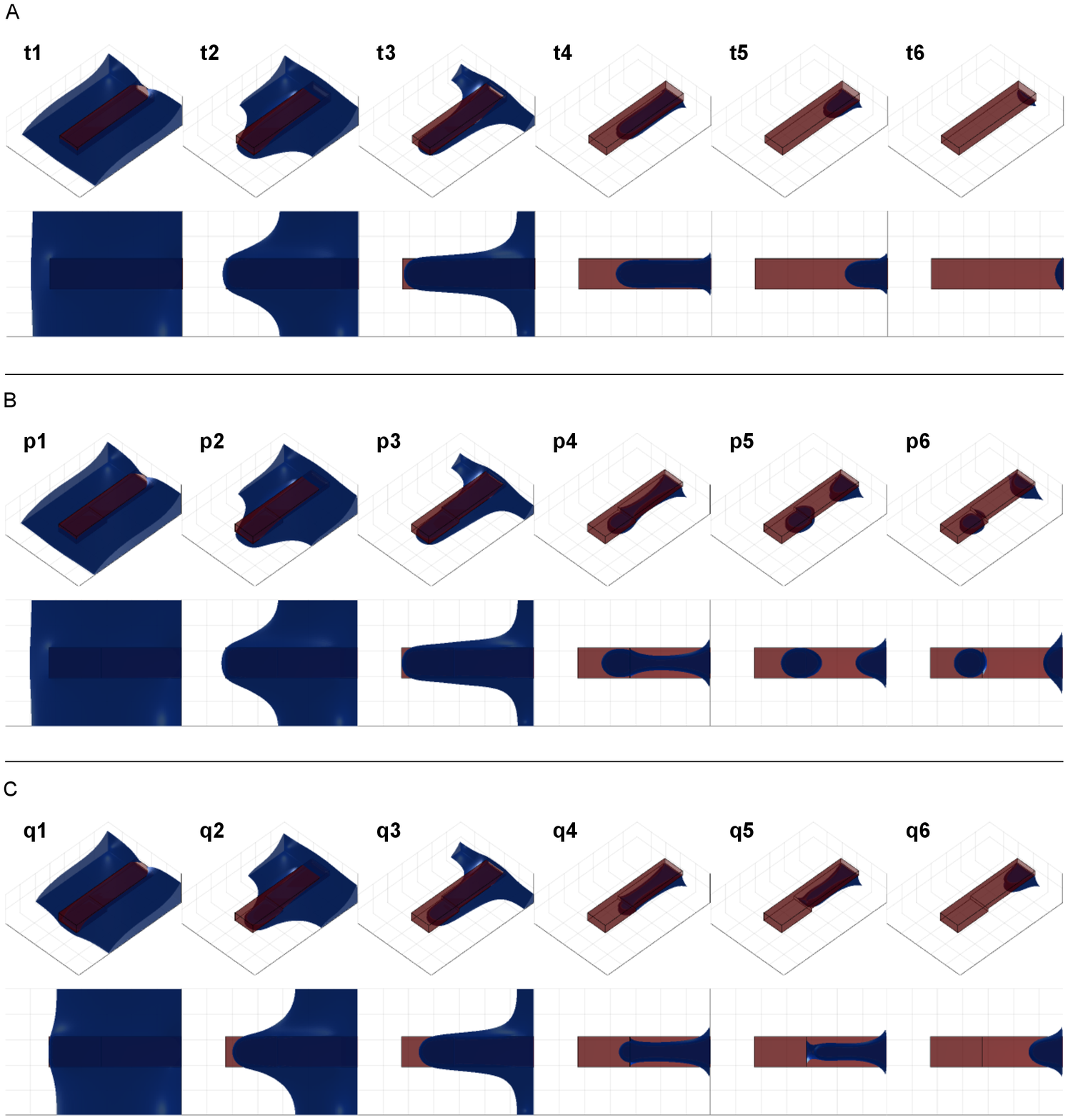} 
\caption{Simulation results showing a front evaporating on rigid hydrophilic cantilevers. 
(a) Evaporation from a uniform (non-stepped) lever. The interface forms a finger-shaped meniscus beneath the cantilever, which recedes until all the liquid evaporates. 
(b) Evaporation on a stepped lever. The base of the lever has a wider gap, compared to the narrower portion, by a fraction $\Delta h/h = 0.4$. The interface forms a spoon shaped finger whose neck collapses. A capillary bridge forms as a result, and moves towards the narrower portion of the lever.  
(c) Effect of evaporation rate. Increasing the evaporation rate by $\sim30\%$ leads to a suppression of the formation of the capillary bridge, even in the presence of a step in the cantilever.
Times are measured in units of the evaporation timescale $t_0=l/E_0$.
The top and bottom rows in each panel correspond to top and bottom views of the lever. 
\label{fig:StraightLever}}
\end{figure*}

Initially, the liquid front covers the entire length of the cantilever (Fig.~\ref{fig:Sketch3D}(b));  evaporation is driven by fixing the local vapour concentration to an out-of-equilibrium value at the open boundary  of the microfluidic cavity (hatched plane marked in Fig.~\ref{fig:Sketch3D}(a)).  
The corresponding imbalance in the chemical potential leads to the retraction of the liquid front.  
Here we use the initial volumetric evaporation rate, $E_0$, as a measure of the strength of evaporation, 
although we note that the instantaneous rate of evaporation will become smaller as the front retreats from the open end of the simulation box. 
We are interested in the interplay between the externally fixed rate of evaporation and the intrinsic timescale of the flow. 
The typical evaporation time is given by $t_0 = l/E_0$.
We anticipate that the timescale of the flow, $t_{\rm flow}=l/u_{\rm cap}$, is controlled by the capillary speed, which we define 
as $u_{\rm cap} \equiv \gamma / 12 \eta_{\rm t}$, where $\gamma$ is the liquid-gas surface tension and $\eta_{\rm t} = \eta_{\rm l} + \eta_{\rm g}$ 
is the sum of the viscosities of the liquid (l) and the gas (g). 
Therefore, the competition between the evaporation and the flow timescales can be quantified by the control parameter 
\begin{equation}
{\rm E} \equiv \frac{E_0}{u_{\rm cap}},
\end{equation} 
which may be thought of as an evaporative capillary number.
In the following, we  report simulation times in units of the evaporation time $t_0$, i.e.~we present results in terms of the dimensionless time $\tilde t = t/t_0$.
 
 \begin{figure*}[t!]
\psfrag{t1}{$\tilde t =0.15$}
\psfrag{t2}{$\tilde t =0.45$}
\psfrag{T1}{$\tilde t =0.44$}
\psfrag{T2}{$\tilde t =0.45$}
\psfrag{L1}[l][l]{(a) Uniform  lever}
\psfrag{L2}[l][l]{(b) Stepped lever}
\psfrag{E}{Evaporation}
\psfrag{F}{Flow}
\includegraphics[width=\textwidth]{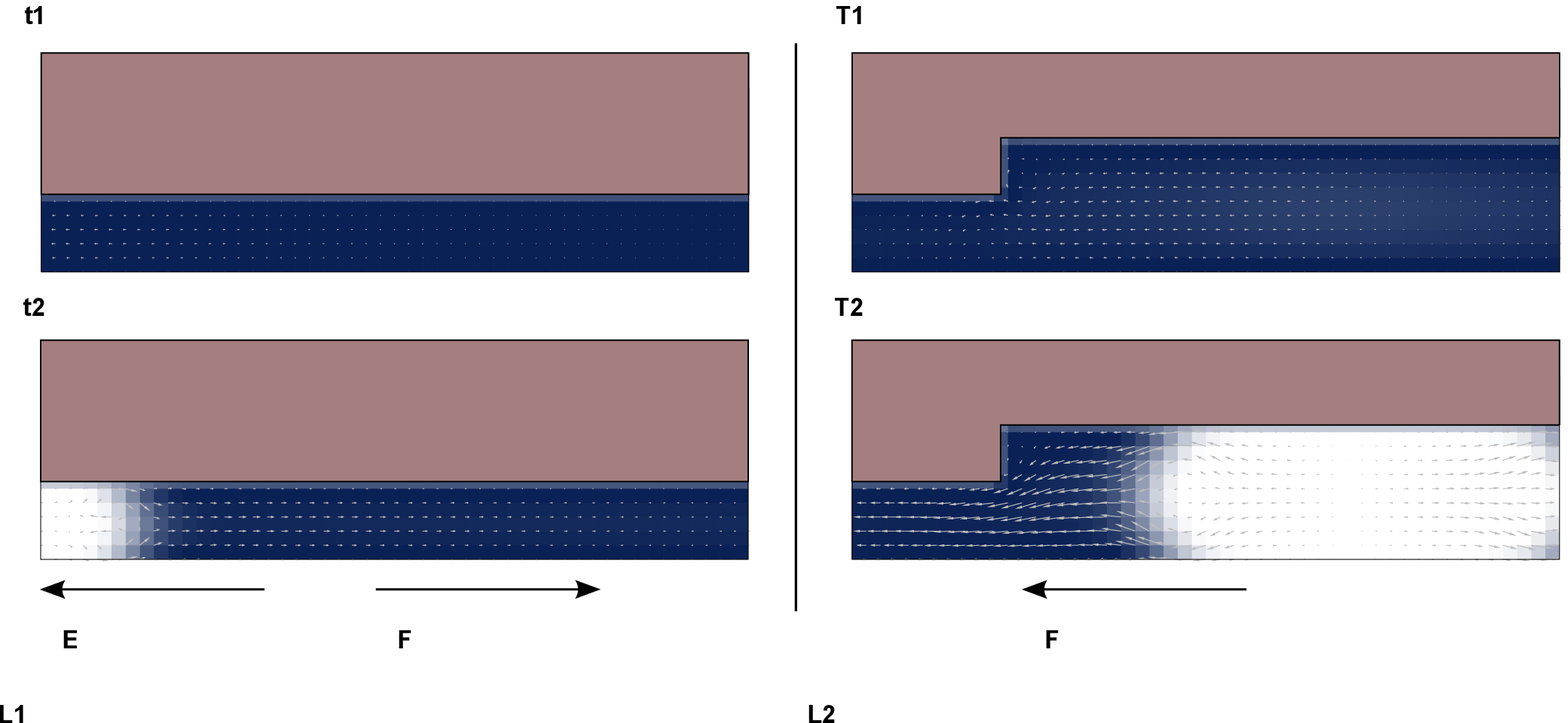}
\caption{\label{fig:StraightLeverV} Flow patterns in the gap below a cantilever upon evaporation at fixed evaporation rate ${\rm E}=0.020$. (a) Uniform cantilever. A flow develops from the tip of the meniscus towards its base. (b) Stepped cantilever with $\Delta h/h=0.4$. The flow pattern points to the narrow portion of the gap, where the Laplace pressure is lower. The scale of the arrows in (a) and (b) is the same, showing that the effect of the step is to generate a strong flow driven by the curvature of the interface.}
\end{figure*}

We first consider the motion of the interface in contact with a straight hydrophilic cantilever parallel to the base. The contact angle of the liquid with the solid is set to $\theta_{\rm e}=30^\circ$ (so that the cantilever is hydrophilic) and the gap with the bottom of the channel has a uniform thickness, $h$.
Fig.~\ref{fig:StraightLever}(a) shows the sequence of configurations adopted by the interface upon evaporation with ${\rm E} =0.02$. 
Initially, the liquid retracts uniformly, until the front comes into contact with the cantilever. 
Thereafter, the covered portion of the liquid recedes more slowly than the exposed surface, so that the interface adopts a finger-like shape beneath the lever. 
This change in conformation is natural since, beneath the cantilever, the liquid has a reduced surface area (so evaporation should proceed more slowly), as well as having a reduced hydrodynamic mobility (due to the friction with the solid walls).
Eventually, though, the finger retracts until the liquid has completely evaporated by a time $\tilde t_{\rm E} \approx 0.58$.

To better understand the effect of confinement, we repeat the simulation using a stepped cantilever so that the gap between the cantilever and the floor of the cavity is $h+\Delta h$ at the lever's base, but $h$ at its tip (see Fig.~\ref{fig:Sketch3D}). 
Figure~\ref{fig:StraightLever}(b) shows a simulation sequence with $\Delta h/h = 0.4$.  
In the early stages, to $\tilde t \simeq 0.29$, the evaporation process follows very closely that observed with the uniform lever. Beyond this point, however, the tip of the interface recedes  more slowly than was observed with a uniform cantilever (compare, for example, the interface configurations at $\tilde t=0.44$ in Figs.~\ref{fig:StraightLever}(a) and~\ref{fig:StraightLever}(b)). 

The difference in the later stages of evaporation between the stepped and straight channel cannot be attributed to either the difference in mobilities or the exposed surface areas (since the stepping is imposed by maintaining the minimum channel thickness these two effects would enhance, rather than diminish, the rate of evaporative motion). Instead, the effective tapering of the channel draws liquid away from the base of the lever towards the tip; this flow replenishes the liquid at the tip of the finger at the expense of the liquid in the centre, which develops a spoon shape, with a thinning neck. 
This neck  eventually collapses, causing a capillary bridge to be snapped off, located roughly at the step where the thickness of the cantilever changes. Once the pinch-off occurs the finger quickly retracts. We note that there is a qualitative similarity between the experiments of \citet{Abe1995} and our simulations (compare Fig.~\ref{fig:StraightLever}(a),(b) with Fig.~\ref{fig:Abe}(b),(a), respectively).

\begin{figure*}[t]
\centering
\psfrag{D1}[l][l]{(a) $\Delta h/h = 0.2,~{\rm E}=0.02$}
\psfrag{D2}[l][l]{(b) $\Delta h/h = 0.6,~{\rm E}=0.02$}
\psfrag{t1}[l][l]{$\tilde t = 0.44$}
\psfrag{t2}[l][l]{$\tilde t = 0.47$}
\psfrag{t3}[l][l]{$\tilde t = 0.50$}
\psfrag{t4}[l][l]{$\tilde t = 0.42$}
\psfrag{t5}[l][l]{$\tilde t = 0.44$}
\psfrag{t6}[l][l]{$\tilde t = 0.45$}
\includegraphics[width=\textwidth]{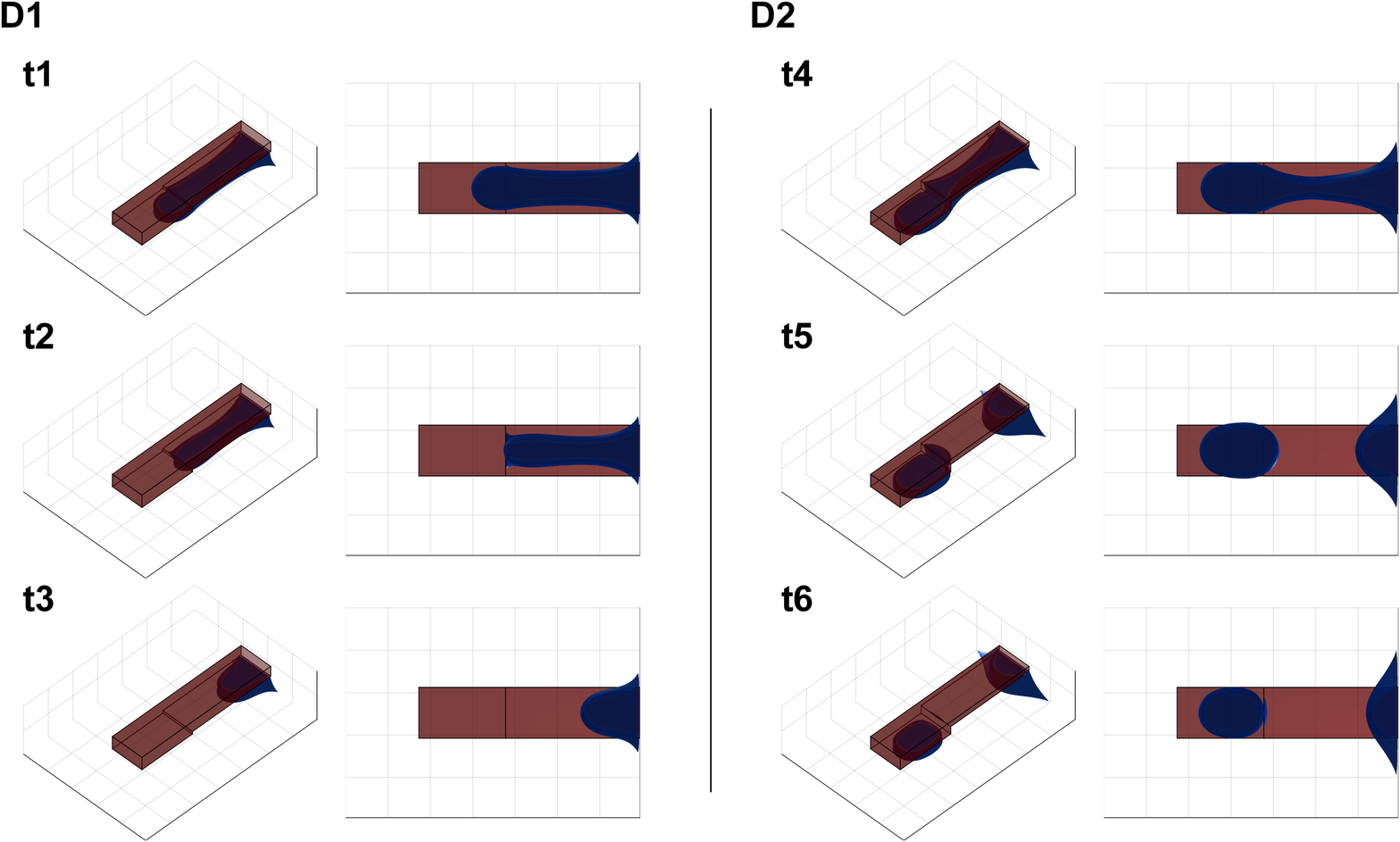} 
\caption{Effect of cantilever geometry on droplet pinch-off. (a) For small step sizes, the interface slows down at the narrow portion of the gap between the lever and the floor or the micro cavity. (b) Above the pinch-off threshold, 
the capillary flow becomes stronger as the step becomes larger, leading to the formation of comparatively larger droplets that appear at earlier times during the evaporation process. Times are measured in units of the evaporation timescale $l/E_0.$
\label{fig:StepLever_h2h6}}
\end{figure*}

\begin{figure*}[t!]
\centering
\psfrag{t1}[r][r]{$\tilde t = 0.10$}
\psfrag{t2}[r][r]{$\tilde t = 0.12$}
\psfrag{t3}[r][r]{$\tilde t = 0.15$}
\psfrag{t4}[r][r]{$\tilde t = 0.18$}
\psfrag{t5}[r][r]{$\tilde t = 0.22$}
\psfrag{t6}[r][r]{$\tilde t = 0.26$}
\includegraphics[width=\textwidth]{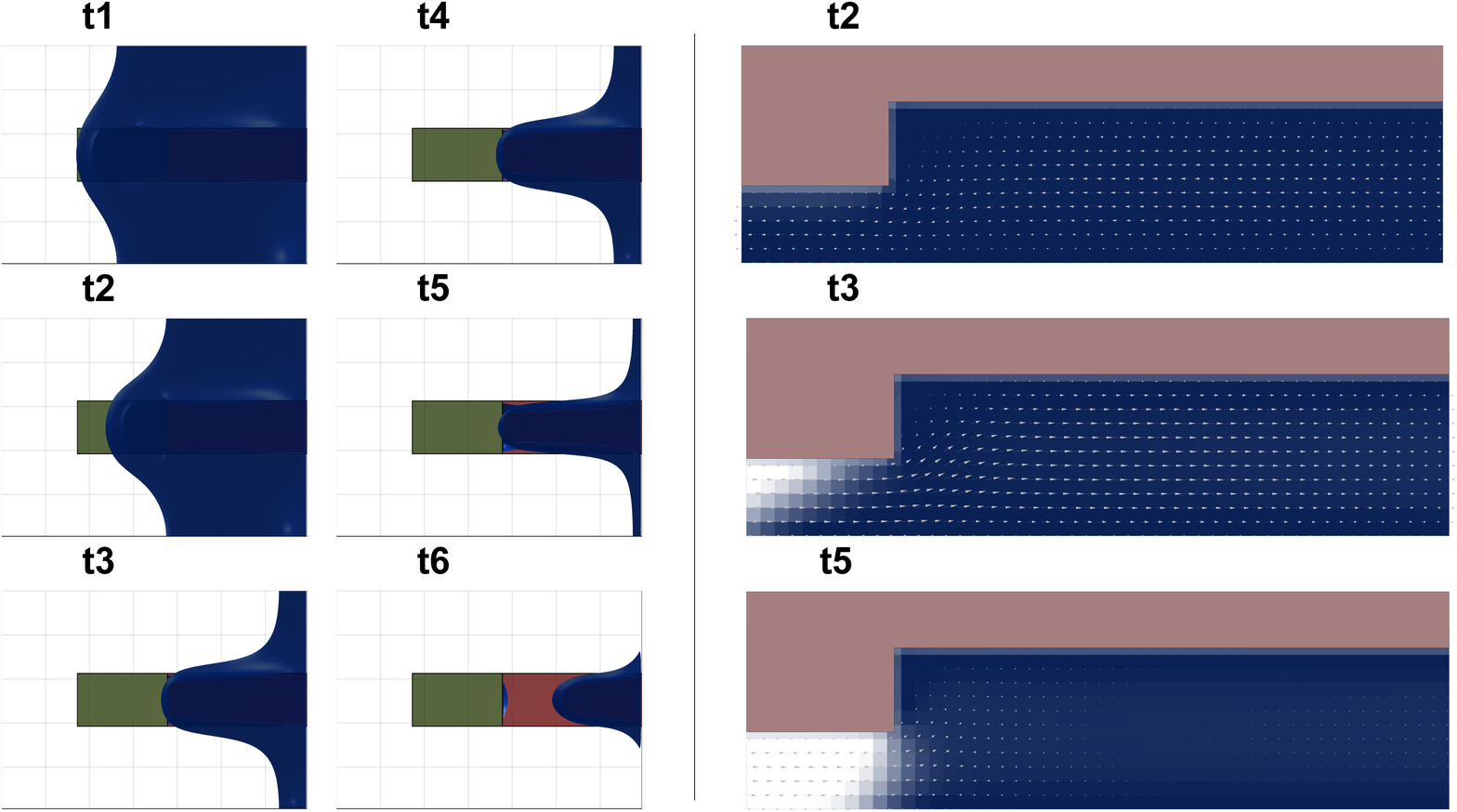} 
\caption{Simulation results of a front evaporating on a rigid stepped cantilever patterned with a hydrophobic patch (green). 
The chemical patterning induces a rise in the capillary pressure between the tip and the body of the meniscus, which creates a flow that counteracts the effect of the channel-width change (right-hand side panels).
As a result, the interface quickly recedes from the hydrophobic patch, preventing the formation of a droplet in the thicker portion of the channel (see bottom panels on the left-hand side).
The dimensionless evaporation rate is ${\rm E}=0.02$.
Times are measured in units of the evaporation timescale $l/E_0.$
\label{fig:StepLeverChem}
}
\end{figure*} 

Despite the slightly slower initial dynamics, the overall effect of the pinch-off is a shorter evaporation time, $\tilde t_{\rm E} \approx 0.54$.  Therefore, the interface configuration, mediated by the solid geometry, plays a crucial role in the evaporation dynamics. 

It is often assumed that the effect of the evaporation rate on the dynamics of a liquid front is purely kinematic, and that modifying this rate merely guides the interface through a set of configurations at different speeds. 
One might therefore expect that changing the evaporation rate ${\rm E}$ would lead to the system proceeding through precisely the same sequence of interface configurations at the same instants of the dimensionless time $\tilde t$ (since our non-dimensionalization of time removes the dependence on ${\rm E}$). 
Surprisingly, our simulations show that this is not the case. Fig.~\ref{fig:StraightLever}(c) shows simulations with the same geometrical parameters as in Fig.~\ref{fig:StraightLever}(b) (i.e.~with $\Delta h/h=0.4$)  but with a  $\sim30\%$ increase in the evaporation, ${\rm E}=0.027$. 
We observe that with this larger evaporation rate, the interface responds by receding from the cantilever without any pinch-off. Closer inspection of the plan-view panels in Fig.~\ref{fig:StraightLever}(c) shows that the tip of the interface recedes significantly faster here than with a lower evaporation rate:  evaporation is stronger at the tip of the finger, preventing the pinch-off and formation of a capillary bridge. The observation that the increase in evaporation rate alters the fluid dynamics in a way that is not accounted for by the rescaling of time is confirmed since the total time taken for evaporation, $\tilde t_{\rm E}({\rm E}=0.027) \approx 0.33$, is smaller than $\tilde t_{\rm E}({\rm E}=0.02) \approx 0.44$.
The change in interface morphology and the evaporation time both show that the effect of increasing the evaporation rate is not purely kinematic. 

To gain further insight into the interplay between evaporation and the fluid dynamics, we compare the flow patterns of the liquid upon evaporation in uniform and stepped lever geometries. 
Figure \ref{fig:StraightLeverV}(a) shows a cross section of the velocity field within a constant gap (corresponding to a uniform cantilever) during evaporation. 
At early times, while the interface is uniform, the liquid remains stationary relative to the walls. 
Then, as the liquid forms a finger, a weak flow develops, pulling liquid from the meniscus towards the end of the channel. 
This effect can be explained in terms of an excess in the Laplace pressure at the tip of the finger relative to its base, which arises from the variation 
of the curvature of the interface in a plane parallel to the bottom of the channel (see bottom views of Fig.~\ref{fig:StraightLever}(a)). 
Such an effect leads to a capillary flow that always assists the retraction of the finger.

For a stepped lever we observe a transient flow in the opposite direction: from the base to the tip of the meniscus (Fig.~\ref{fig:StraightLeverV}(b)). 
This reversed flow is due to a (relatively) large difference in the transverse curvature of the interface between the tip and base of the lever, 
which causes the capillary flow to point \emph{in} the direction of confinement \cite{Reyssat2014,Gorce2016}. 
This flow also drives the pinch-off of the liquid bridge, which occurs on a much faster timescale, $\Delta \tilde t \simeq 0.45 -0.44  = 0.01$, than evaporation, $\tilde t_{\rm E} \sim 1$.  
We also note that immediately following pinch-off ($\tilde t = 0.45$), the capillary bridge has a visible curvature difference between its front and back edges. 
The corresponding difference in the Laplace pressure results in a strong capillary flow, shown in Fig.~\ref{fig:StraightLeverV}(b), which causes the bridge to move towards the tip of the lever at $\tilde t =0.47$. 
In turn, the in-plane curvature of the remnant finger leads to a faster retraction of the liquid towards the base of the lever (not shown in the figure).

The net capillary flow that results from the competition between the in-plane and transverse interface curvatures determines whether the pinch-off of a capillary bridge occurs. 
For example, setting $\Delta h/h=0.2$ results in no pinch-off at all, while fixing $\Delta h/h=0.6$ leads to a pinch-off event at a slightly earlier time ($\tilde t\approx 0.43$) 
and to a relatively larger capillary bridge compared to that observed for $\Delta h / h = 0.4$ (see Fig.~\ref{fig:StepLever_h2h6}). 

Further control of this instability can be obtained using the wettability of the lever (which has the effect of reversing the transverse curvature of the front). Fig.~\ref{fig:StepLeverChem}~shows a simulation sequence corresponding to $\Delta h/h=0.6$, but with the narrow part of the lever hydrophobic  (equilibrium  contact angle $\theta_{\rm e}=150^\circ$).  
The resulting gradient in wettability now opposes the geometrical effect of the tapering of the gap, and results in a smaller meniscus curvature in the narrow portion of the channel. 
Therefore, the meniscus retreats very quickly from the narrow portion of the channel and only slows down when it reaches the wider portion of the channel: 
pinch-off is averted by the use of hydrophobic patterning.

\subsection{Analytical model}

\label{sec:2DTheory}

Our simulation results have identified three ingredients that control the stability of an evaporating liquid-gas interface in a confined geometry: variations in channel height, wettability and the rate of evaporation. 
{ To understand the interaction of these effects quantitatively, we now present a model of the dynamics of a meniscus in confinement
that captures the effect of changes in wettability, confinement and evaporation rate at the interface on the stability of the front.
}

\begin{figure}[t!]

\centering

\psfrag{x}{$x$}
\psfrag{y}{$y$}
\psfrag{z}{$z$}
\psfrag{L}{$L$}
\psfrag{W}{$W$}
\psfrag{th}{\textcolor{white}{$\theta(\xm)$}}
\psfrag{hx}{\textcolor{white}{$h(x)$}}
\psfrag{ly}{\textcolor{white}{$(\xm(y,t),y)$}}
\psfrag{c0}{$\mu_0$}
\psfrag{csat}[r][r]{$\mu=\mu_{\rm e}$}
\psfrag{cxy}{$ \mu(x,y)$}
\psfrag{pgxy}{$p_{\rm g}(x,y)$}
\psfrag{plxy}{\textcolor{white}{$p_{\rm l}(x,y)$}}
\includegraphics[width=\columnwidth]{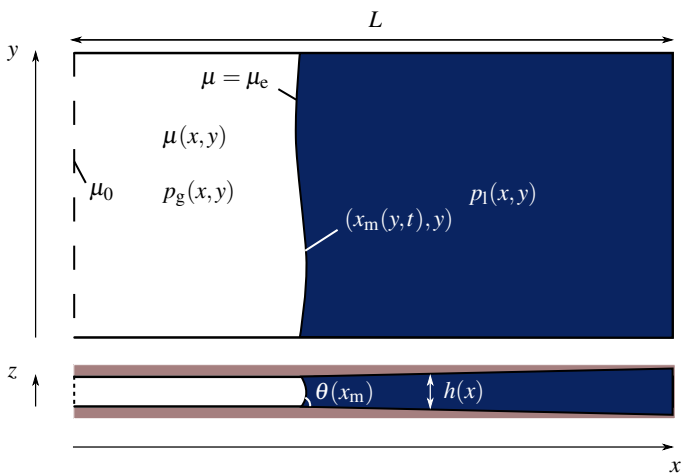}
\caption{\label{fig:2DSketch} Schematic of a liquid undergoing evaporation in a channel of length $L$, infinite width, and varying thickness $h(x)$. The evaporation is driven 
by fixing the chemical potential of vapour molecules to a constant value $\mu_0$ at the left end of the system. 
The interface, indicated by the solid curve, has a transverse curvature (in the $xz$ plane) which
varies according to the local channel thickness, $h(x)$, and the local wetting angle, $\theta(\xm)$.}
\end{figure}

To facilitate the analytical solution of such a model, we consider a simplified geometry, in which an initially flat liquid front evaporates in 
a channel of length $L$ and infinite width (Fig.~\ref{fig:2DSketch}). 
The liquid is confined by rigid, impermeable walls at the top, bottom and back.  
The liquid meets a gas phase at an interface, whose in-plane coordinates, averaged over the thickness of the channel, are $x=\xm(y,t)$ for $-\infty<y<\infty$.
The gas phase is connected to an ambient reservoir through an open end at the front of the channel, located at $x=0$. 
The gap between the bounding walls, $h(x)$, is allowed to vary in the $x$ direction. 
The resulting tapering mimics the non-uniform gap between the cantilever and the floor of the channel in our LB simulations. 
We also include a local dependence of the contact angle of the meniscus with the channel walls, $\theta(\xm)$,
which models the effect of a change in wettability along the surface of the cantilever.
The closed end of the channel ensures that a static uniform interface configuration exists in the absence of evaporation. 

The evaporation of the front is driven by an imbalance in the chemical potential between the liquid and the gas. 
We focus on the limit of small channel aspect ratios, $h/L\ll1$, and slow gap variations, $|h'|\ll1$. 
In this limit, the chemical potential, $\mu$, only depends on the in-plane coordinates $x$ and $y$ to leading order (variations in the $z$-direction across the gap are assumed negligible due to the effect of mass diffusion).
Therefore, an imbalance in $\mu$ can be modelled by using the boundary conditions 
$ \mu (0,y)=\mu_0,$ and $\mu (\xm,y)=\mu_{\rm e},$ where $\mu_0$ and $\mu_{\rm e}$ correspond to the chemical potential of the gas reservoir and the liquid, respectively. 
In the absence of convection, and for slow evaporation relative to the typical diffusive timescale of the vapour in the gas, the chemical potential 
relaxes to a quasi-static configuration which satisfies Laplace's equation,  
\begin{equation}
\label{eq:Diffusion}
\frac{\partial^2 \mu }{\partial x^2 } + \frac{\partial^2 \mu}{\partial y^2 }= 0. 
\end{equation} 
Once a solution for the chemical potential is found, the rate of evaporation can be calculated by evaluating the normal mass flux at the interface, i.e., $E =-(M/\Delta\phi) {\nabla} \mu\cdot {\bf \hat n}|_{x=\xm,y}$, where $M$ is the diffusivity, $\Delta \phi$ is the difference in the relative concentration of liquid molecules between the liquid and the gas and ${\bf \hat n}$ is the 
unit normal to the interface. 

For a flat interface, $\xm(y,t) = \xm^0(t)$, the chemical potential is a function of $x$ alone, and hence depends linearly on $x$, $\mu = \mu^0(x,t)$. 
Therefore, the evaporation rate,   
\begin{equation}
E_0 = - \frac{M}{\Delta \phi}\left.\frac{{\rm d} \mu^0}{{\rm d}x}\right|_{x=\xm,y}=
- \frac{M}{\Delta \phi}\frac{\mu_{\rm e}-\mu_0}{\xm^0},
\end{equation}
determines the instantaneous speed of the interface, via the kinematic condition
\begin{equation}
\frac{{\rm d} \xm^0}{{\rm d} t } =  - E_0.
\end{equation}
Because the chemical potential difference is kept fixed while the interface moves away from the reservoir, the interface motion is purely diffusive, $\xm^0(t)\sim t^{1/2}.$ Furthermore, 
the liquid and gas phases remain at rest, with uniform pressure in each phase (though these pressures differ by a constant as required by the meniscus curvature).

When the planar interface is perturbed, the interface develops an in-plane curvature; 
for a gently perturbed front, this curvature may be written
\begin{equation}
C_{\parallel}(\xm(y,t)) \approx -  \frac{\partial^2 \xm }{\partial y^2}.
\end{equation} 
Because of the non-uniform channel thickness and wettability, the transverse interfacial curvature may also vary: assuming the meniscus locally remains a circular arc, geometry gives that 
\begin{equation}
C_{\perp} (\xm(y,t))=\frac{2 \cos\bigl[\theta (\xm) - h'(\xm)/2\bigr]}{h(\xm)}.
\label{eq:curvature}
\end{equation}

The effect of the interface perturbation is to create a non-uniform evaporation rate, 
\begin{equation}
E = - \frac{M}{\Delta \phi }\left . \frac{\partial \mu}{\partial x }\right|_{x=\xm,y},
\end{equation} 
but also results in a variation of the pressure along the interface, due to Laplace's law, 
\begin{equation}
p_{\rm l} (\xm,y) = p_{\rm g} (\xm,y)  - \gamma \left(C_\parallel + C_{\perp} \right),
\label{eq:YoungLaplace}
\end{equation}
where $p_{\rm l}$ and $p_{\rm g}$ are the pressure fields in the liquid and the gas, and $\gamma$ is the liquid-gas surface tension.
These perturbations will decay far away from the interface, i.e., $\partial_xp_{\rm g} = \partial_x p_{\rm l} = 0$ as $|x-\xm|\rightarrow \infty$, 
but close to the front the local pressure gradient will result in a capillary flow.
Within our approximation, this flow can be described using Darcy's Law:
\begin{equation}
{\bf v}_i  = -\frac{h^2}{12 \eta_i} {\nabla} p_i,
\end{equation}
where ${\bf v}_i$ is the velocity field of the $i$th phase and $\eta_i$ the corresponding dynamic viscosity. 
Imposing incompressibility, ${\nabla} \cdot {\bf v}_i$ =0, the fluid dynamics reduces 
to a partial differential equation for the pressure field in each phase, namely
\begin{equation}
h^3 \left( \frac{\partial^2 p_i}{\partial x^2} + \frac{\partial^2 p_i}{\partial y^2} \right) 
+ 3 h^2 h' \frac{\partial p_i}{\partial x} = 0. 
\label{eq:pressure}
\end{equation}

The kinematic condition on the interface is also modified (since the liquid next to the interface is moving too); we have that
\begin{equation}
\frac{ \partial \xm}{\partial t} = - E+ u_{\rm h},
\label{eq:InterfaceSpeed} 
\end{equation}
where the hydrodynamic velocity, 
\begin{equation}
u_{\rm h} = -\frac{h^2}{12\eta_{i}} \left.\frac{\partial p_{i}}{\partial x}\right|_{x=\xm,y}.
\end{equation}

To analyse the stability of the planar meniscus configuration, we consider small sinusoidal  perturbations to the interface, i.e.  
\begin{equation}
\xm(y,t) =  \xm^0(t) + \epsilon \sin(k y)\exp(\sigma t), 
\end{equation}
where $\epsilon/\xm^0 \ll 1 $ is the amplitude of the perturbation,  $k$ is its wave-number and $\sigma$ its growth rate. We also choose 
the linear wettability and tapering profiles  $ \theta(x) = \theta_{\rm e} + \alpha (x-x_0)$  and $h(x) =  h_0 + \beta (x-x_0),$ where $h_0$, $\theta_{\rm e}$ 
and $x_0$ are reference values, and $\alpha$ and $\beta$ are the relevant rates of change with the $x$ coordinate.

Expanding $\mu$ and $p_i$ in powers of $\epsilon$ about the flat-interface solution and using Eqs.~(\ref{eq:YoungLaplace}) and~(\ref{eq:InterfaceSpeed}), we find that the growth rate of the perturbation  is given by
\begin{equation}
\sigma (k) = \left[
\frac{\gamma h_0^2}{12 \eta_{\rm t}} \left( \frac{2 \alpha \sin\theta_{\rm e}}{h_0} + \frac{2 \beta\cos \theta_{\rm e}}{h_0^2}
- k^2 \right) 
- E_0  \right] k. 
\label{eq:sigmak}
\end{equation}

Before discussing this result further, it is useful to express it in dimensionless form.  
Using the capillary speed, $u_{\rm cap} = \gamma/12\eta_{\rm t}$, and the channel thickness, $h_0$, as the characteristic 
units for speed and length, we obtain 
\begin{equation}
\hat{\sigma}(\hat{k}) =  ({\rm Ch}  + {\rm Ta}- {\rm E} )\hat{k} - \hat{k}^3,
\label{eq:2DDispersionRelation}
\end{equation}   
where we have defined $\hat{\sigma} \equiv  h_0 \sigma/u_{\rm cap}$, $\hat{k} \equiv k h_0$. 

Apart from the dimensionless evaporation rate, ${\rm E} = E_0/u_{\rm cap}$, 
two additional dimensionless parameters emerge: 
\begin{equation}
{\rm Ch}  \equiv 2 h_0 \alpha \sin \theta_{\rm e},\quad {\rm Ta} \equiv 2 \beta\cos \theta_{\rm e} ,
\end{equation} which we term the chemical number and tapering number, respectively. 
The key emergent feature of this result is that, although the evaporation rate depends on time, it does so on a slow timescale, 
which allows for the different terms in the dispersion relation to compete with each other.

Equation~(\ref{eq:2DDispersionRelation}) contains a purely stabilizing term $\sim - \hat k^3$ arizing from the relaxation of the in-plane curvature of the front. 
The remaining terms, each proportional to $\hat k$, may be stabilizing or destabilizing, depending on the sign of the dimensionless parameters ${\rm Ch}$, ${\rm Ta}$ and ${\rm E}$. Note that for ${\rm E} > 0$, corresponding to evaporation, the interface tends to be stabilized: perturbations to the interface that bring it closer to the open boundary also increase the evaporative flux locally, ironing out those perturbations (and similarly for perturbations that take the interface further away). It is also worth noting that setting ${\rm E} <0$, corresponding to condensation, has a destabilizing effect. 
The chemical number and the tapering number behave in similar ways (and so may be combined additively):  ${\rm Ch}>0$ (i.e.~ $ \alpha>0$) 
corresponds to a situation in which the liquid occupies the less wettable parts of the channel and the chemical gradient will act to pull liquid into the more wettable parts while ${\rm Ta}>0$ (e.g.~$\theta_{\rm e}<\pi/2$ and $\beta>0$) corresponds to a wetting liquid, which tends to invade narrower portions of the channel~\cite{Keiser2016}. 
Both ${\rm Ch}>0$ and ${\rm Ta}>0$ therefore destabilize the front, while switching their signs renders them stabilizing.

\begin{figure*}[t!]
\psfrag{E}{E}
\psfrag{x}[c][c]{${\rm Ta+Ch-E}$}
\psfrag{s}{S}
\psfrag{u}{U}
\psfrag{0.01}{0.01}
\psfrag{0.02}{0.02}
\psfrag{0.03}{0.03}
\psfrag{0}{0}
\psfrag{-0.5}{-0.5}
\psfrag{-0.4}{-0.4}
\psfrag{-0.3}{-0.3}
\psfrag{-0.2}{-0.2}
\psfrag{-0.1}{-0.1}
\psfrag{0.1}{0.1}
\psfrag{0.2}{0.2}
\includegraphics[width=\textwidth]{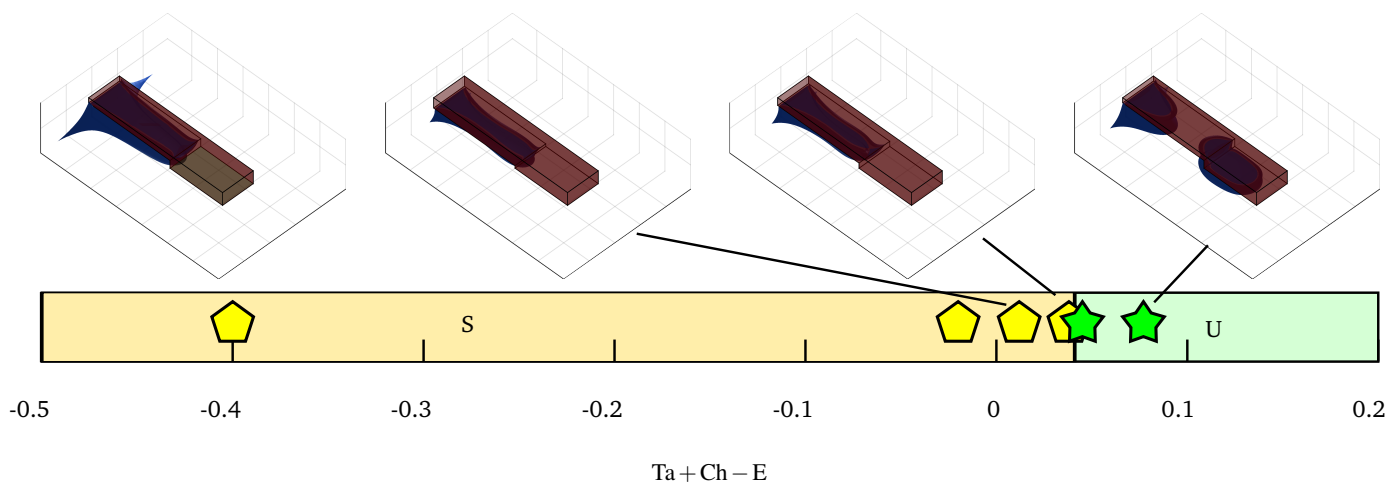}
\caption{\label{fig:PhaseMap} Stability phase map for a liquid front evaporating in a cantilever geometry. The horizontal axis compares the effect of evaporation to variations in the lever wettability and local thickness. Sufficiently positive ${\rm Ta} + {\rm Ch} -{\rm E}$ tends to destabilize the interface, leading to the pinch-off of capillary bridges (stars). Varying Ta, Ch, and E  can be used to suppress the instability (pentagons). The background shading indicates the regions of stable (S) and unstable (U) fronts observed in the simulations.}
\end{figure*}

The interplay between these three effects is expected to lead to an unstable front whenever
\begin{equation}
\label{eq:2DPhaseDiagram}
{\rm Ch} + {\rm Ta} - {\rm E} > 0.
\end{equation}
In such a case, perturbations of the interface of wave number $\hat k$ are expected to be unstable provided that
\begin{equation}
\label{eq:2DCriticalk}
\hat k<\hat k_{\rm c} = \left ({\rm Ch} + {\rm Ta} -{\rm E}\right)^{1/2}
\end{equation} (i.e.~surface tension stabilizes very short wavelengths, but perturbations with sufficiently long wavelengths are unstable). We also note that the fastest growing mode  has $\hat k_{\rm m} = \hat k_{\rm c}/\sqrt{3}$ and hence that we expect to observe a wavelength
\begin{equation}
\lambda_{\rm m}=\frac{2\sqrt{3}\pi}{\left ({\rm Ch} + {\rm Ta} -{\rm E}\right)^{1/2}}.
\end{equation}

Whilst these results are strictly only valid for a channel of uniform transverse cross-section in the $xz$ plane (see Fig.~\ref{fig:2DSketch}), we expect that the criterion given by Eq.~(\ref{eq:2DPhaseDiagram}) remains qualitatively valid 
for more complicated geometries, including the cantilever setup explored in our simulations.
To calculate the corresponding chemical and tapering numbers we use the approximations $\alpha \approx \Delta \theta_{\rm e}/l$ and $\beta \approx \Delta h /l$, 
where $\Delta \theta_{\rm e}$ and $\Delta h$ are the changes in contact angle and height of the cantilever, and $l$ its length. 
Fig.~\ref{fig:PhaseMap} shows a regime diagram that summarizes the results of our Lattice-Boltzmann simulations. 
For small evaporation rates, ${\rm E} \ll1$, the front destabilizes as Ta increases, leading to the formation of capillary bridges. 
However, an otherwise unstable front can be stabilized either by introducing a gradient in the channel wettability (a finite, negative Ch) or by increasing the evaporation rate. 
%

%%%%%%%%%%%%%%%%%%%%%%%%%%%%%%%%%%%%%%%%%%%%%%%%%%%%%%%%%%%%%%%%%%%%%%%%%%%

\section{Discussion}

The results presented in this paper illustrate general features governing the stability of liquid fronts undergoing a phase change in a  geometry relevant for MEMS and other micro-scale systems.
While the motion and stability of the liquid front is affected by gradients in the Laplace pressure along the interface induced by the geometry of the MEMS or its wettability, 
our results show that evaporation can qualitatively change the dynamics (controlling the pinch-off of capillary bridges in our specific cantilever setup). 
This effect is relevant even at a relatively small evaporation rate, compared to the typical speed of the hydrodynamic capillary flow.

The main physical mechanism behind this behaviour can be understood in terms of our mathematical model, which provides a stability condition that compares the interplay between the dimensionless parameters Ch, Ta, and E (representing the effects of varying wetting properties, geometry and a phase change, respectively).  
In particular, since the stability or otherwise of a planar interface is determined only by the condition ${\rm Ch + Ta - E} > 0$ (Eq.~\eqref{eq:2DPhaseDiagram}), we see that it is the relative sizes of the channel patterning and tapering, relative to the evaporation rate, that matters.

These ideas can be used to understand the dynamics of fronts undergoing a phase change in more complicated geometries, and hence as a first design principle to either avoid or to control the motion and stability of liquid fronts in micro-scale geometries. 
In particular, we note that, unlike in previous work~\cite{Hadjittofis2016}, in our formulation it  is not necessary to have large dimensionless evaporation rates for the stabilizing effect of the evaporation rate to become important: the key is for the evaporation rate to be strong enough to counteract the destabilizing effects of the tapering and wettability gradients discussed here.
This may go some way to resolving the apparent paradox that evaporation cannot happen quickly enough to effect the dynamics of instability, and yet has apparently been observed to do so.  
Finally, we note that the importance of the parameter ${\rm Ch + Ta - E}$ is consistent with observations of the effect of evaporation and tapering in MEMS fabrication~\cite{Abe1995,Abe1996}. 
In the fabrication of rigid micro-beams the liquid is observed to evaporate without a droplet pinching off. This corresponds to the case  ${\rm Ch + Ta }=0$, which we expect to be stable since ${\rm Ch + Ta - E}<0$. However, with a flexible beam (so that ${\rm Ta}>0$) the evaporation is seen to lead to droplet pinch-off if the beam is long enough (which we interpret as corresponding to a sufficiently large degree of tapering) \cite{Abe1996}. Similarly, Abe \emph{et al.}~\cite{Abe1995} noted that increasing the temperature led to less stiction in a way that could not be accounted for by the expected decrease in surface tension coefficient. 
Instead, we  suggest that this result may be due to a change in sign of ${\rm Ch + Ta - E}$ as the evaporation rate is increased so that evaporation suppresses pinch-off. However, this suggestion requires direct experimental tests. We hope that the insights offered in this paper, together with recent studies of how interfacial morphology affects evaporation rates \cite{Saenz2017}, will motivate further experimental studies to quantify and understand the interface dynamics during evaporation and condensation in confinement.

%\bibliography{bibliography}
%\bibliographystyle{rsc} %the RSC's .bst file

\providecommand*{\mcitethebibliography}{\thebibliography}
\csname @ifundefined\endcsname{endmcitethebibliography}
{\let\endmcitethebibliography\endthebibliography}{}

\end{document}